\documentclass[conference]{IEEEtran}
\IEEEoverridecommandlockouts
\usepackage{cite}
\usepackage{amsmath,amssymb,amsfonts}
\usepackage{graphicx}
\usepackage{textcomp}
\usepackage{xcolor}
\usepackage{booktabs}
\usepackage{array}
\usepackage{url}

\def\BibTeX{{\rm B\kern-.05em{\sc i\kern-.025em b}\kern-.08em
    T\kern-.1667em\lower.7ex\hbox{E}\kern-.125emX}}

\begin{document}

\title{ADVENT: LLM-Driven Automatic Predicate Invention for ILP}

\author{%
\resizebox{0.96\textwidth}{!}{%
\begin{tabular}{@{}c@{\hspace{0.02\textwidth}}c@{\hspace{0.02\textwidth}}c@{}}
\begin{tabular}[t]{c}
Tingting Yu\\
\textit{Department of Information Management}\\
\textit{National Sun Yat-Sen University}\\
Kaohsiung, Taiwan\\
tingyui0213@gmail.com
\end{tabular}
&
\begin{tabular}[t]{c}
Pei-Cing Huang\\
\textit{Department of Information Management}\\
\textit{National Sun Yat-Sen University}\\
Kaohsiung, Taiwan\\
pcpeicing@gmail.com
\end{tabular}
&
\begin{tabular}[t]{c}
Chan Hsu\\
\textit{Department of Information Management}\\
\textit{National Sun Yat-Sen University}\\
Kaohsiung, Taiwan\\
chanshsu@gmail.com
\end{tabular}
\\[1.6ex]
\multicolumn{3}{c}{%
\begin{tabular}{@{}c@{\hspace{0.10\textwidth}}c@{}}
\begin{tabular}[t]{c}
Chan-Tung Ku\\
\textit{Department of Information Management}\\
\textit{National Sun Yat-Sen University}\\
Kaohsiung, Taiwan\\
kuchantung@gmail.com
\end{tabular}
&
\begin{tabular}[t]{c}
Yihuang Kang\\
\textit{Department of Information Management}\\
\textit{National Sun Yat-Sen University}\\
Kaohsiung, Taiwan\\
ykang@mis.nsysu.edu.tw
\end{tabular}
\end{tabular}}
\end{tabular}
}
}

\maketitle

\begin{abstract}
Predicate invention (PI), the creation of new predicates to extend the hypothesis space, remains a critical bottleneck in Inductive Logic Programming (ILP). Existing methods rely on domain expertise and produce semantically opaque predicates, hindering adaptation to unfamiliar domains and cross-task reuse. We present ADVENT, an LLM-driven PI mechanism for ILP. ADVENT pairs LLM abductive generation with Prolog deductive verification, forming an iterative loop in which concrete execution results guide the LLM to refine candidate predicates. The mechanism leverages Large Language Models to identify implicit patterns in structured relational data and invent auxiliary predicates with meaningful names and definitions. Invented predicates and learned rules accumulate in a knowledge pool for cross-task reuse. Experiments on nine poker-hand concepts across seven LLMs show that LLM-driven PI achieves 58\% success rate where ILP alone fails entirely, formal verification raises this to 80\%, and the knowledge pool yields gains up to +31 percentage points, while producing human-interpretable rules. These results suggest that ADVENT offers a promising direction for automating predicate invention and enabling cross-task knowledge reuse in ILP.
\end{abstract}


\section{Introduction}

Many real-world concepts are inherently relational: a molecule is mutagenic not because of any single atom, but because of how atoms are bonded together. Inductive Logic Programming (ILP)~\cite{cropper2022ilp30} is one of the few machine learning paradigms capable of learning such relational concepts~\cite{gabbay2008logical} as human-interpretable logical rules. However, ILP's performance depends heavily on predefined background knowledge (BK)~\cite{cropper2022ilp30}, a set of predicates that encode known facts and relational structures about the domain, analogous to features in traditional ML. Without suitable predicates, the system cannot discover rules that explain the target concept well. Predicate invention (PI) addresses this limitation by introducing auxiliary predicates that extend the hypothesis space beyond the original BK, making it one of the most critical open challenges in ILP~\cite{cropper2022ilp30}. Yet existing PI approaches suffer from two fundamental limitations. First, without human-provided specifications to guide the search, PI devolves into exhaustive syntactic enumeration over an explosive space, making it difficult to automatically discover helpful patterns~\cite{cropper2022ilp30,cropper2021popper}. Second, existing PI methods operate through purely syntactic symbol manipulation and lack conceptual understanding of what they invent~\cite{gentili2025predicate}. Unnamed predicates such as \textit{inv\_1} and \textit{inv\_2} have no notion of the patterns these symbols encode, making it hard to assess which prior inventions are relevant to a new task. Furthermore, predicates that build upon earlier inventions become increasingly difficult to interpret. Together, these limitations restrict ILP's ability to learn in new domains and hinder its potential for lifelong learning~\cite{cropper2020forgetting}.

\begin{figure*}[!t]
\centering
\includegraphics[width=0.85\textwidth, height=5.2cm, keepaspectratio=false]{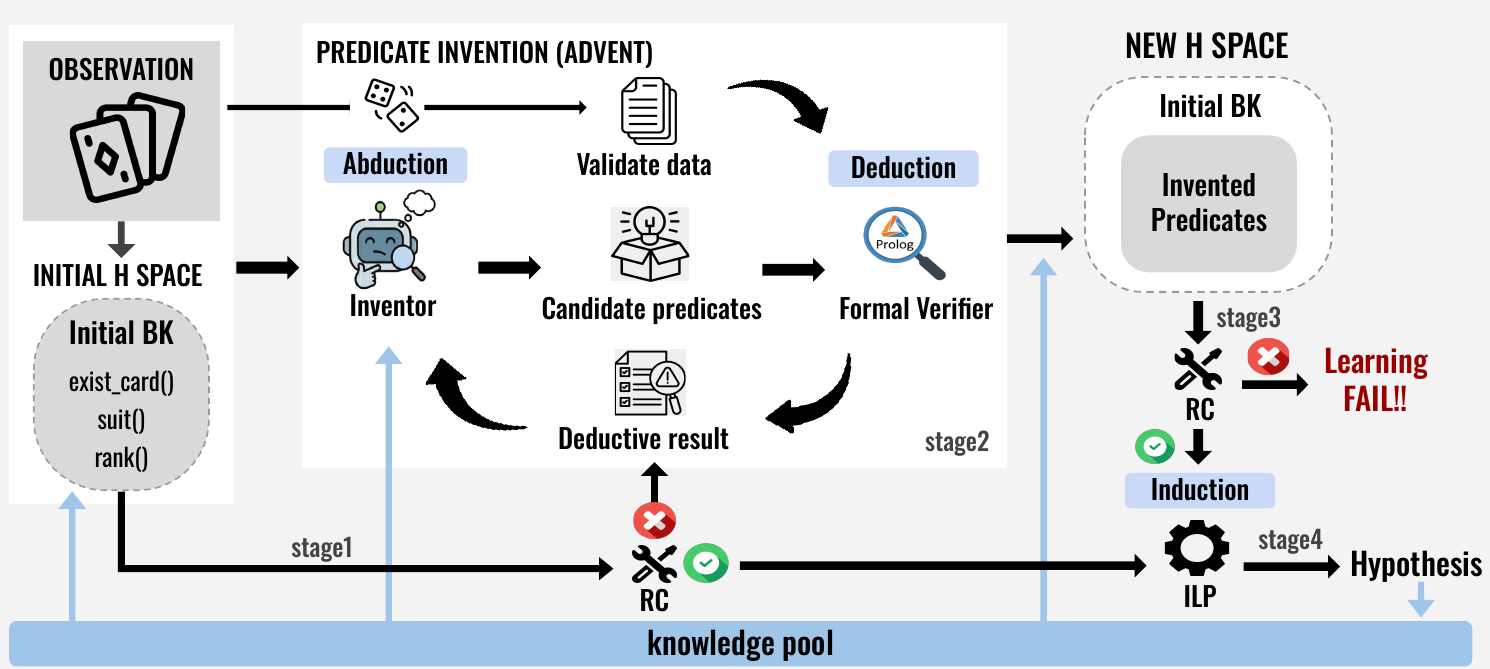}
\caption{Overall pipeline.}
\label{fig:pipeline}
\end{figure*}

LLMs offer a promising path to address these limitations. Trained on large-scale corpora, LLMs encode broad prior knowledge and demonstrate strong capabilities in code generation, abductive reasoning~\cite{chen2021evaluating,peng2025abductive}, and logic programs synthesis~\cite{gandarela2025inductive},  suggesting they could inspect structured data, recognize implicit patterns, and operate directly within ILP's symbolic formalism, opening new possibilities for automating predicate invention. Moreover, because LLMs generate predicates with natural language understanding, the invented predicates inherently carry meaningful names (e.g., \textit{same\_color}) and interpretable definitions, providing a basis for cross-task relevance assessment.

However, relying solely on LLM reasoning introduces reliability concerns~\cite{ji2023hallucination}. In preliminary experiments, we observe two recurring types of errors: syntactic violations of logic programming conventions, and predicates that capture overly specific patterns that fail to generalize. Recent studies have shown that iteratively refining LLM outputs with formal verification can substantially improve their reasoning quality~\cite{sun2024clover,singh2026verge}. In the context of ILP, Prolog~\cite{korner2022prolog}, the predominant logic programming language used in ILP, offers a deductive engine that serves as a natural verifier: by executing candidate predicates on subsets of examples, it precisely reveals whether an invented predicate generalizes as intended. These observations motivate three research questions:

\begin{itemize}
\item RQ1. Can LLMs serve as an effective mechanism for predicate invention in ILP?
\item RQ2. Can formal deductive verification effectively guide LLM predicate refinement and improve ILP learning performance?
\item RQ3. Can LLM-driven PI support cross-task predicate reuse and composition while preserving interpretability?
\end{itemize}

To address these questions, we present ADVENT, an LLM-driven PI mechanism that integrates LLM abductive reasoning with Prolog deductive verifier, and combines them with ILP inductive learning. Invented predicates and learned hypotheses accumulate in a knowledge pool, enabling reuse of prior knowledge across tasks. Our main contributions are as follows:

\begin{itemize}
\item We propose ADVENT, an LLM-driven PI mechanism that identifies implicit patterns in structured data and generates corresponding predicates for ILP.
\item We integrate a Prolog-based formal verifier that provides concrete deductive results to guide predicate refinement.
\item We show that LLMs can interpret invented predicate definitions and assess their relevance to new tasks, enabling selective reuse and composition that supports cumulative learning with interpretable rules.
\end{itemize}

\section{Background}

We briefly review ILP and relevant prior work to situate ADVENT within the existing literature. An ILP task takes three inputs: background knowledge (BK), which encodes known facts and relational structures about the domain; positive examples ($E^+$); and negative examples ($E^-$) of the target concept. The goal is to find a hypothesis ($H$), a set of first-order logic rules expressed through predicates, variables, and quantifiers, that maximizes coverage of $E^+$ while minimizing coverage of $E^-$. This representation is expressive, logically verifiable, and human-interpretable, making ILP well-suited to scientific discovery and safety-critical domains~\cite{cropper2022ilp30}. However, PI within this formalism has been a long-standing challenge~\cite{cropper2022ilp30}. Meta-Interpretive Learning (MIL) systems such as Metagol~\cite{muggleton2015meta} employ metarules as templates to guide the invention of new predicates, enabling the construction of recursive and compositional hypotheses. However, the expressiveness of Metagol is bounded by the metarules provided, requiring domain expertise to design appropriate templates. Popper~\cite{cropper2021popper} takes a different approach by using hypothesis constraints derived from failed hypotheses to prune the search space, supporting PI through mode declarations that specify the permitted predicate signatures. While more flexible than template-based methods, Popper still relies on manually defined mode declarations to control the search.

These limitations have motivated recent work on integrating LLMs into the ILP pipeline. Yang et al.~\cite{yang2026hypothesis} proposed a multi-agent LLM framework to automatically build BK and language bias from natural-language descriptions of data attributes, substantially reducing manual BK design. However, the generated predicates correspond only to explicit surface-level patterns; implicit relations requiring cross-example reasoning remain outside their scope. Our work targets such implicit patterns, using LLMs not to extract what is described but to invent what is not directly observable. While LLMs can generate logic programs, their outputs in formal domains remain prone to errors~\cite{pan2023logiclm}. Recent work has shown that iteratively verifying and refining LLM outputs with formal tools can substantially improve correctness~\cite{sun2024clover,singh2026verge}. We apply this paradigm using Prolog's deductive engine: candidate predicates are executed against training examples, and grounded results guide iterative refinement.

Beyond improving invention, a practical PI mechanism must support knowledge reusability across tasks. ILP's compositional structure naturally supports this: hypotheses learned from earlier tasks can serve as BK for subsequent ones~\cite{cropper2022ilp30}. However, as BK grows, the system faces the relevance problem~\cite{cropper2022ilp30}: without a mechanism to judge which prior knowledge applies to a new task, accumulation leads to catastrophic remembering~\cite{cropper2020forgetting}, where irrelevant BK degrades both search efficiency and learning performance. Forgetgol~\cite{cropper2020forgetting} mitigates this through selective forgetting based on usage frequency, but this remains a structural heuristic that does not address the root cause. Our approach addresses the problem at its source: by generating predicates with meaningful names and interpretable definitions, LLM-driven PI produces knowledge that the LLM can later assess for relevance to new tasks.

\section{Methodology}

We propose ADVENT, a predicate invention mechanism that pairs LLM-based abductive generation with Prolog-based deductive verifier. This mechanism is embedded in a neuro-symbolic pipeline that additionally employs the ILP system for rule induction, mirroring Peirce's triadic reasoning cycle of abduction, deduction, and induction~\cite{he2025reasoning}. Fig.~\ref{fig:pipeline} illustrates the overall pipeline, which operates in four stages for each learning task.

\begin{itemize}
\item \textit{Stage~1: Initial Hypothesis Space Representation Check.} At the onset of learning, the framework first evaluates whether the initial BK already provides sufficient representational power. We introduce a lightweight heuristic, termed the Representation Check (RC), to make this determination (detailed in Section~III-A). If RC passes, the framework skips predicate invention and proceeds directly to Stage~4.

\item \textit{Stage~2: LLM-based Predicate Invention Loop.} When RC fails, it signals that the available BK lacks the expressiveness to capture the underlying patterns, and the framework enters an iterative invention loop. In this loop, the LLM automatically synthesizes new auxiliary predicates to enrich the hypothesis space, while Prolog provides deductive verification to guide refinement.

\item \textit{Stage~3: Post-invention Representation Check.} After PI, the framework re-applies RC over the expanded hypothesis space to validate the effectiveness of the invented predicates. If RC passes, the framework proceeds to rule induction; otherwise, the case is recorded as a learning failure.

\item \textit{Stage~4: ILP Rule Induction.} With an adequate hypothesis space confirmed, the ILP learner induces a symbolic hypothesis that generalizes across the provided examples.
\end{itemize}

In addition to this per-task pipeline, a cross-task mechanism accumulates knowledge across learning episodes. Invented predicates that contribute to successful induction, along with learned rules, accumulate in a knowledge pool that expands the initial hypothesis space of new targets and provides the LLM with a reference for reuse or composition during PI. The following sections detail each component in turn.

\begin{figure}[!b]
\centering
\includegraphics[width=0.85\columnwidth]{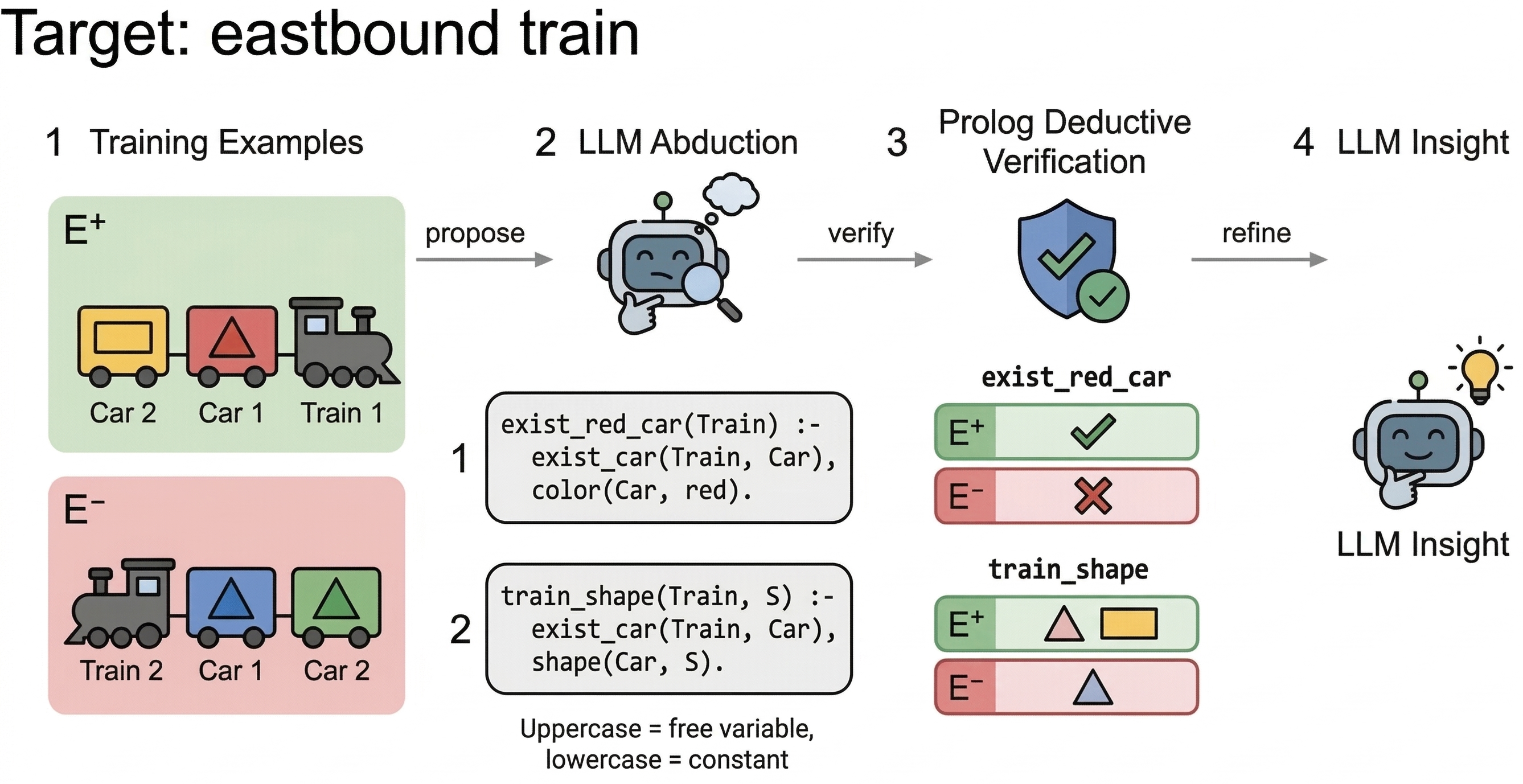}
\caption{Illustration of formal deductive verification.}
\label{fig:verification}
\end{figure}

\subsection{Representation Check (RC)}

To determine the necessity of PI, we introduce RC, a lightweight ILP with restricted variable-level search. Variable-level search replaces specific attribute values with shared variables, forcing the learner to capture relational patterns across objects. For example, in Fig.~\ref{fig:verification}, the most specific rule for train~2 would be $\mathit{shape}(Car1, S) \wedge \mathit{shape}(Car2, S)$, where the shared variable $S$ captures ``same shape'' without committing to a specific value. RC operates in three steps: (1) 10\% of the positive and negative examples are sampled; (2) 1\% of the positive examples from this subset are selected as seeds, each serving as the starting point for an independent rule search ; (3) if all trials yield only a ground unit clause that covers solely its seed example with no other positive examples, RC signals that the current BK lacks the predicates needed to express the underlying relational patterns and triggers PI. The 10\% example sample and 1\% seed selection were chosen as lightweight heuristics to keep RC inexpensive while still checking whether any seed yields a non-trivial variable-level generalization.

\subsection{LLM-PI Loop: Iterative Predicate Invention with Deductive Verification}

Once the framework determines via RC that the initial BK is insufficient, this module implements the core predicate invention cycle, in which the LLM abductively proposes candidate predicates and Prolog deductively verifies them to support closed-loop refinement.

\subsubsection{LLM Predicate Generation}
At each iteration, the LLM receives a structured prompt consisting of two components: (a) the current task's BK schema along with a sample of positive and negative examples, and (b) the full inventory of the knowledge pool, listing previously invented predicates with their names and definitions. Given this context, the LLM is prompted to identify implicit patterns across examples and propose one or more candidate predicates, each with a Prolog definition, its arity, and usage specifications such as argument types and input/output modes for integration into the ILP induction system. These predicates may compose directly from the initial BK or, by assessing relevance from their names and definitions, build upon prior inventions to form higher-level abstractions.

\subsubsection{Formal Deductive Verification}
After the LLM proposes candidate predicates, Prolog serves as a deductive verifier through two layers of validation. The first layer performs syntactic and structural checks: Prolog attempts to consult the predicate definition, catching runtime errors, unsatisfiable clauses, and undefined predicates. Candidates that fail receive the specific error message for immediate correction. The second layer evaluates predicate behavior on examples. A subset of examples is sampled from both $E^+$ and $E^-$, and each candidate is executed via Prolog's deductive engine. As illustrated in Fig.~\ref{fig:verification}, a unary predicate such as $\mathit{exist\_red\_car}(Train)$ produces a Boolean result per example, while a higher-arity predicate such as $\mathit{train\_shape}(Train, S)$ returns satisfying value tuples. In both cases, results are partitioned by class ($E^+$ vs.\ $E^-$) and presented contrastively, enabling the LLM to assess whether the pattern is discriminative or requires refinement. A maximum iteration limit is enforced to prevent excessive computation.

\subsection{ILP Rule Induction}

Once RC confirms that generalizable patterns exist, the framework proceeds to rule induction. The auxiliary predicates used to describe these patterns are injected into the original BK to form the refined search space. Since RC has already verified that relational patterns exist at the variable level, the ILP learner then performs a constant-level search, grounding variables to specific values. For example, the same train~2 in Fig.~\ref{fig:verification} would produce $\mathit{shape}(Car1, triangle) \wedge \mathit{shape}(Car2, triangle)$. This allows the learner to capture instance-level distinctions that variable-level rules abstract away, complementing the LLM's high-level inventions with fine-grained, data-driven rules.

\section{Experiment}

\subsection{Dataset and Task Transfer}

We evaluate the proposed framework on a relational concept learning task using the Poker Hand dataset from the UCI Machine Learning Repository~\cite{cattral2002poker}. Each instance is a five-card hand with two attributes per card (suit and rank), labeled by a category such as one pair, flush, or straight. To prevent the LLM from exploiting prior knowledge of poker rules, we transfer the task into a classic ILP benchmark, Michalski Train problem~\cite{michalski1983theory}: each hand is mapped to a train, each card to a car, suit to shape, rank to load number, and each poker-hand category to a distinct train direction (e.g., one pair as eastbound, two pairs as westbound), preserving the relational structure while making the concepts unfamiliar to the LLM. The initial BK consists of three base predicates: $\mathit{exist\_car/2}$, $\mathit{shape/2}$, and $\mathit{load\_num/2}$. For each target concept, negative examples are drawn from all lower-ranked hands. To verify this transfer, we prompted the LLM to identify the scenario behind the data; the model consistently interpreted it as a Michalski's Trains puzzle with no recognition of the poker domain. Conversely, replacing suit and rank predicates with generic variable names such as $X_1$ and $X_2$ led the model to immediately recognize the poker scenario, confirming that the transfer effectively abstracts away domain-specific cues.

\subsection{Experimental Setup}

We adopt Aleph~\cite{aleph2026} as the ILP engine, a top-down search system without native PI support. The ILP-only baseline induces over the base BK under two modes: position-dependent, a flattened representation where each card occupies a fixed slot, and position-free, a fully relational representation where cards are accessed through logical relations without fixed positions, better reflecting real poker scenarios. Three additional settings combine LLM-driven PI with Aleph under position-free mode, differing in verification: no verification (single-pass invention), LLM self-critique, and formal verification (ADVENT). Models were accessed through OpenRouter using provider-specific identifiers, e.g., \textit{openai/gpt-5-codex} and \textit{x-ai/grok-4.1-fast}. All three operate under a sequential learning protocol with the knowledge pool, where concepts are learned sequentially from one pair to royal flush in ascending complexity. To isolate the effect of knowledge accumulation, we include a formal verification setting without the knowledge pool. Each LLM-based setting is repeated for five runs across nine concepts (45 trials per model). During each PI iteration, the LLM receives five positive and two negative examples per concept as input, with at most three iterations per trial. A trial is counted as successful only when the framework induces the correct ground-truth rule.
\begin{table}[!t]
\caption{Learning success rates across experimental conditions.\textsuperscript{\dag}}
\label{tab:main_results}
\centering
\scriptsize
\setlength{\tabcolsep}{2pt}
\begin{tabular}{l c c c c}
\toprule
\multicolumn{3}{c}{ILP-only position-dependent}& 6/9 & \\
\multicolumn{3}{c}{ILP-only position-free}& 0/9& \\
\midrule
\multicolumn{5}{c}{\textbf{LLM PI} (position-free) \textbf{+ ILP}}\\
\midrule
\textbf{Model} & \textbf{No Verif.} & \textbf{LLM Crit.} & \textbf{ADVENT} & \textbf{ADVENT(--KP)} \\
\midrule
GPT-5-Codex & \textbf{84\% [71--92\%]} & 80\% [66--89\%] & 82\% [69--91\%] & 71\% [57--82\%] \\
grok-4.1-fast & 64\% [50--77\%] & \textbf{78\% [64--87\%]} & \textbf{78\% [64--87\%]} & 73\% [59--84\%] \\
gemini-2.5-pro & 36\% [23--50\%] & \textbf{80\% [66--89\%]} & 76\% [61--86\%] & 71\% [57--82\%] \\
qwq-32b & 62\% [48--75\%] & 49\% [35--63\%] & \textbf{67\% [52--79\%]} & 53\% [39--67\%] \\
GLM-5 & 78\% [64--87\%] & 73\% [59--84\%] & \textbf{80\% [66--89\%]} & \textbf{80\% [66--89\%]} \\
MiMo-V2-Omni & 31\% [20--46\%] & 80\% [66--89\%] & \textbf{93\% [82--98\%]} & 56\% [41--69\%] \\
deepseek-r1 & 51\% [37--65\%] & 73\% [59--84\%] & \textbf{87\% [74--94\%]} & 64\% [50--77\%] \\
\midrule
\textbf{Avg.} & \textbf{58\% [53--63\%]} & \textbf{73\% [68--78\%]} & \textbf{80\% [76--84\%]} & \textbf{67\% [62--72\%]} \\
\bottomrule
\end{tabular}

{\scriptsize \textsuperscript{\dag}Wilson score intervals are computed over 45 trials (5 $\times$ 9 concepts) per cell; since concepts vary in difficulty, they characterize aggregate performance stability rather than a single underlying rate.}
\end{table}
\section{Results and Discussion}

\subsection{Main Results}

As Table~\ref{tab:main_results} shows, ILP alone learns at most 6 out of 9 concepts under position-dependent binding, and fails entirely without it (0/9). In contrast, even without verification, LLM-driven PI under position-free mode achieves an average success rate of 58\% [53--63\%], demonstrating that LLMs can identify implicit patterns and invent useful predicates that substantially extend ILP's representational capacity (RQ1). Adding formal verification (ADVENT) raises the average to 80\% [76--84\%], whose confidence interval does not overlap with the no-verification condition, confirming that Prolog deductive verification effectively guides predicate refinement (RQ2). By comparison, LLM self-critique achieves 73\% [68--78\%], suggesting that grounded execution results provide more reliable refinement signals than LLM self-assessment alone. Furthermore, removing the knowledge pool reduces the average from 80\% to 67\% [62--72\%], with drops in six out of seven models and non-overlapping confidence intervals, indicating that the LLM's ability to reuse previously invented predicates substantially benefits learning (RQ3). The following subsections examine each research question in detail. For readability, all rules and predicates in the following section are presented using the original poker terminology.

\subsection{LLM-Driven Predicate Invention}

Table~\ref{tab:per_concept} presents per-concept learning results for ILP alone and ADVENT+ILP, and Table~\ref{tab:learned_rules} further illustrates this gap through representative examples. The two ILP-only modes differ in how cards are accessed. Position-dependent mode uses predicate $\mathit{has\_card}(+Hand, -C1, -C2, -C3, -C4, -C5)$, where $C1$ through $C5$ correspond to fixed card slots. Position-free mode uses $\mathit{exist\_card}(+Hand, -C)$, which retrieves an arbitrary card without assuming any ordering. Consider one pair, which requires two cards of the same rank. In position-dependent mode, ILP alone must enumerate all $\binom{5}{2} = 10$ slot combinations, producing a verbose rule set. In position-free mode, ILP alone fails entirely: comparing two cards drawn via $\mathit{exist\_card}$ requires an inequality constraint $C1 \neq C2$ to ensure distinct card references, yet ILP alone cannot introduce such constraints autonomously.

The LLM, by contrast, naturally incorporates it when inventing $\mathit{has\_duplicate\_rank}$, thereby yielding the compact rule: $\mathit{one\_pair}(A)$ :- $\mathit{has\_duplicate\_rank}(A)$. The gap widens with straight, which requires five cards in consecutive ranks. Detecting this pattern demands cross-object arithmetic operations that ILP's refinement operators cannot construct without pre-defined building blocks. The LLM addresses this by composing three auxiliary predicates $\mathit{min\_rank}$, $\mathit{max\_rank}$, and $\mathit{has\_consecutive\_ranks}$, while ILP's subsequent search discovers an additional rule covering the A-10-J-Q-K edge case, illustrating a complementary division of labor. These examples reveal two categories of representational gaps that LLM-driven PI bridges: implicit relational constraints (such as variable inequality) and arithmetic relations (such as consecutive sequences), both fundamentally inaccessible to syntax-driven search.

\begin{table}[!t]
\caption{Per-concept learning results.}
\label{tab:per_concept}
\centering
\footnotesize
\setlength{\tabcolsep}{2pt}
\begin{tabular}{l c c c c c c c c c}
\toprule
& 1P & 2P & 3K & ST & FL & FH & 4K & SF & RF \\
\midrule
ILP-only (pos.) & \checkmark & \checkmark & \checkmark & $\times$ & \checkmark & \checkmark & \checkmark & $\times$ & $\times$ \\
ILP-only (free.) & $\times$ & $\times$ & $\times$ & $\times$ & $\times$ & $\times$ & $\times$ & $\times$ & $\times$ \\
\midrule
ADVENT+ILP (+KP) & 63 & 71 & 91 & 69 & 91 & 89 & 91 & 94 & 63 \\
ADVENT+ILP (--KP) & 57 & 69 & 71 & 60 & 100 & 66 & 83 & 63 & 34 \\
\bottomrule
\multicolumn{10}{l}{\scriptsize pos. = position-dependent, free. = position-free; values in \%.}\\
\multicolumn{10}{l}{\scriptsize 1P = one pair, 2P = two pairs, 3K = three of a kind;}\\
\multicolumn{10}{l}{\scriptsize ST = straight, FL = flush, FH = full house, 4K = four of a kind;}\\
\multicolumn{10}{l}{\scriptsize SF = straight flush, RF = royal flush.}
\end{tabular}
\end{table}

\subsection{Effectiveness of Formal Deductive Verification}
Having characterized the patterns LLMs can identify, we turn to the role of formal verification in refining these inventions. As Table~\ref{tab:main_results} shows, formal verification improves success rates for six out of seven models. The benefit is most pronounced for models with lower baselines: MiMo rises from 31\% to 93\% and DeepSeek from 51\% to 87\%, with non-overlapping confidence intervals in both cases. Models that already generate near-optimal predicates in a single pass (GPT-5-Codex, GLM-5) show marginal gains due to ceiling effects. These results confirm that Prolog deductive verification effectively guides predicate refinement and improves ILP learning performance.
\subsection{Formal Verification vs.\ LLM Self-Critique}

\begin{figure*}[!t]
\centering
\includegraphics[width=0.88\textwidth, height=6.0cm, keepaspectratio=false]{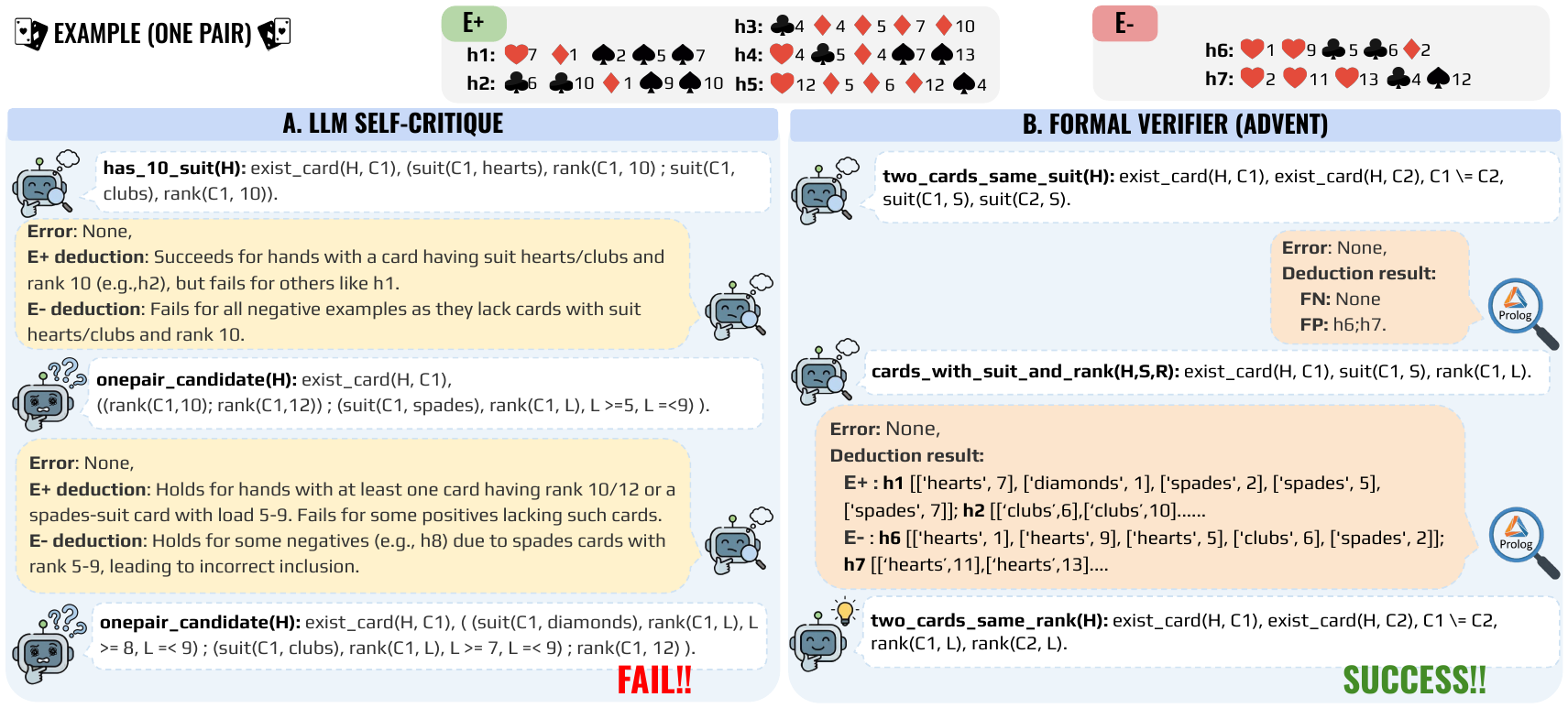}
\caption{Case study comparing formal verification and LLM self-critique on the one pair task.}
\label{fig:case_study}
\end{figure*}

We compare formal verification against LLM self-critique as an alternative refinement signal. While LLM self-critique also improves over no verification on average, it is less reliable: QWQ-32B drops from 62\% to 49\%, suggesting that unreliable self-assessment can mislead refinement. Formal verification outperforms LLM self-critique in five of seven models, with the exception of Gemini-2.5-Pro, where the model's strong self-assessment partially compensates for its weak baseline. Fig.~\ref{fig:case_study} illustrates this difference on the one pair task. Under formal verification, the LLM initially proposes a predicate checking shared suit, but Prolog's per-example results immediately flag false positives among negative examples. In the next iteration, a higher-arity predicate retrieves suit-rank pairs for each example, and the contrastive layout makes the duplicate-rank pattern in positive examples directly visible. Under LLM self-critique, by contrast, the model lacks this per-example granularity and progressively overfits to superficial features across iterations, ultimately failing to capture the underlying pattern. This highlights that evaluating predicate behavior on concrete data is inherently a deduction task, and Prolog provides this with formal correctness.

\begin{table}[!t]
\caption{Learned rules for one pair and straight.}
\label{tab:learned_rules}
\centering
\footnotesize
\setlength{\tabcolsep}{3pt}
\renewcommand{\arraystretch}{1.0}
\newcolumntype{C}[1]{>{\centering\arraybackslash}m{#1}}
\newcolumntype{R}[1]{>{\raggedright\arraybackslash}m{#1}}
\begin{tabular}{C{0.4cm} C{0.5cm} R{0.7\columnwidth}}
\toprule
\multicolumn{3}{c}{\textbf{ILP-only (pos.)}} \\
\midrule
\multicolumn{2}{C{0.9cm}}{1P} & \textbf{one\_pair(Hand)} :- has\_card(Hand,C1,C2,C3,C4,C5), rank(C1,G), rank(C2,G). \newline
\textbf{one\_pair(Hand)} :- has\_card(Hand,C1,C2,C3,C4,C5), rank(C1,G), rank(C3,~G)...... \newline
(C(5,2) = 10 position-specific clauses) \\
\midrule
\multicolumn{3}{c}{\textbf{ADVENT+ILP}} \\
\midrule
1P & PI & \textbf{has\_duplicate\_rank(Hand)} :- exist\_card(Hand,~C1), exist\_card(Hand,~C2), C1~\textbackslash=~C2, rank(C1,~R), rank(C2,~R). \\[12pt]
 & ILP & \textbf{one\_pair(Hand)} :- has\_duplicate\_rank(Hand). \\
\midrule
ST & PI & \textbf{min\_rank(Hand,~Min)} :- exist\_card(Hand,~C1), rank(C1,~Min), \textbackslash+((exist\_card(Hand,~C2), rank(C2,~M), M~$<$~Min)). \newline
\textbf{max\_rank(Hand,~Max)} :- exist\_card(Hand,~C1), rank(C1,~Max), \textbackslash+((exist\_card(Hand,~C2), rank(C2,~M), M~$>$~Max)). \newline
\textbf{has\_consecutive\_ranks(Hand)} :- min\_rank(Hand,~Min), max\_rank(Hand,~Max), Max~=:=~Min~+~4, \textbackslash+~has\_duplicate\_rank(Hand). \\[40pt]
 & ILP & \textbf{straight(Hand)} :- has\_consecutive\_ranks(Hand). \newline
\textbf{straight(Hand)} :- exist\_card(Hand,C1), rank(C1,10), exist\_card(Hand,C2), rank(C2,11), exist\_card(Hand,C3), rank(C3,12), exist\_card(Hand,C4), rank(C4,13), exist\_card(Hand,C5), rank(C5,1). \\
\bottomrule
\end{tabular}
\end{table}

\subsection{Cross-Task Predicate Reuse and Composition}
We now examine whether LLM-driven PI enables effective cross-task reuse. Table~\ref{tab:per_concept} breaks down the impact by concept. The largest gains concentrate in compositional concepts (straight flush +31, royal flush +29, full house +23), which leverage knowledge from earlier learning. However, for simpler tasks such as flush ($-9$), where only a single shared attribute is needed, exposing prior predicates appears to introduce unnecessary complexity. The drop on flush suggests that an unfiltered knowledge pool can introduce distracting predicates for simple concepts, indicating a trade-off between compositional reuse and irrelevant-context interference. As shown in Table~\ref{tab:learned_rules}, the predicate $\mathit{has\_consecutive\_ranks}$, invented during learning straight, reuses the previously invented $\mathit{has\_duplicate\_rank}$ from one pair to exclude hands with repeated ranks. Such cross-task composition arises because the LLM can interpret prior predicate definitions and assess their relevance to the current task. This compositionality extends further: the hypothesis for straight flush composes $\mathit{has\_all\_same\_suit}$ (from flush) and $\mathit{has\_consecutive\_ranks}$ (from straight), yet remains immediately comprehensible, maintaining ILP's core advantage of interpretability.

\section{Conclusion}

This paper presents ADVENT, an LLM-driven automatic predicate invention mechanism for ILP that integrates LLM abductive reasoning with Prolog deductive verification. Experiments demonstrate that LLMs can identify implicit patterns and invent auxiliary predicates that extend ILP's representational capacity, achieving a 58\% success rate where ILP alone fails entirely. Formal deductive verification raises this to 80\%, confirming that grounded execution results effectively guide predicate refinement. The knowledge pool further benefits compositional concepts, with gains up to +31 percentage points, demonstrating that LLM-driven PI supports cross-task reuse and composition while preserving interpretability. However, the current design provides the full pool to the LLM without explicit filtering, relying on the model's ability to interpret predicate definitions and assess their relevance; as the pool grows, this approach is bounded by the LLM's context window, and an increasing number of predicates may dilute the model's attention. In addition, the current evaluation is limited to a transformed poker-hand domain. Future work will develop selective retrieval mechanisms to filter the pool based on relevance and evaluate ADVENT on broader real-world relational datasets.

{\small
\bibliographystyle{IEEEtran}
\bibliography{references}

@article{cropper2022ilp30,
  author    = {Andrew Cropper and Sebastijan Duman\v{c}i\'{c}},
  title     = {Inductive Logic Programming at 30: A New Introduction},
  journal   = {arXiv},
  year      = {2022},
  eprint    = {2008.07912},
  archiveprefix = {arXiv},
  note      = {Accessed: Nov.\ 09, 2024},
  url       = {http://arxiv.org/abs/2008.07912}
}

@book{gabbay2008logical,
  editor    = {Dov M. Gabbay and others},
  title     = {Logical and Relational Learning},
  series    = {Cognitive Technologies},
  publisher = {Springer-Verlag Berlin Heidelberg},
  address   = {Berlin, Heidelberg},
  year      = {2008},
  doi       = {10.1007/978-3-540-68856-3}
}

@article{cropper2021popper,
  author    = {Andrew Cropper and Rolf Morel},
  title     = {Learning Programs by Learning from Failures},
  journal   = {Machine Learning},
  volume    = {110},
  number    = {4},
  pages     = {801--856},
  year      = {2021},
  month     = apr,
  doi       = {10.1007/s10994-020-05934-z}
}

@article{gentili2025predicate,
  author    = {Elisabetta Gentili and Tony Ribeiro and Fabrizio Riguzzi and Katsumi Inoue},
  title     = {Predicate Renaming via Large Language Models},
  journal   = {arXiv},
  year      = {2025},
  doi       = {10.48550/ARXIV.2510.25517}
}

@inproceedings{cropper2020forgetting,
  author    = {Andrew Cropper},
  title     = {Forgetting to Learn Logic Programs},
  booktitle = {Proceedings of the AAAI Conference on Artificial Intelligence},
  volume    = {34},
  number    = {04},
  pages     = {3676--3683},
  year      = {2020},
  month     = apr,
  doi       = {10.1609/aaai.v34i04.5776}
}

@article{chen2021evaluating,
  author    = {Mark Chen and Jerry Tworek and Heewoo Jun and others},
  title     = {Evaluating Large Language Models Trained on Code},
  journal   = {arXiv},
  year      = {2021},
  eprint    = {2107.03374},
  archiveprefix = {arXiv},
  doi       = {10.48550/arXiv.2107.03374}
}

@article{peng2025abductive,
  author    = {Yifei Peng and Yaoli Liu and Enbo Xia and others},
  title     = {Abductive Logical Rule Induction by Bridging Inductive Logic Programming and Multimodal Large Language Models},
  journal   = {arXiv},
  year      = {2025},
  eprint    = {2509.21874},
  archiveprefix = {arXiv},
  doi       = {10.48550/arXiv.2509.21874}
}

@article{gandarela2025inductive,
  author    = {Jo\~{a}o Pedro Gandarela and Danilo S. Carvalho and Andr\'{e} Freitas},
  title     = {Inductive Learning of Logical Theories with {LLMs}: An Expressivity-Graded Analysis},
  journal   = {arXiv},
  year      = {2025},
  eprint    = {2408.16779},
  archiveprefix = {arXiv},
  doi       = {10.48550/arXiv.2408.16779}
}

@article{ji2023hallucination,
  author    = {Ziwei Ji and Nayeon Lee and Rita Frieske and others},
  title     = {Survey of Hallucination in Natural Language Generation},
  journal   = {ACM Computing Surveys},
  volume    = {55},
  number    = {12},
  pages     = {248:1--248:38},
  year      = {2023},
  doi       = {10.1145/3571730}
}

@inproceedings{sun2024clover,
  author    = {Chuyue Sun and Ying Sheng and Oded Padon and Clark Barrett},
  title     = {Clover: Closed-Loop Verifiable Code Generation},
  booktitle = {AI Verification: First International Symposium, SAIV 2024, Montreal, QC, Canada, July 22--23, 2024, Proceedings},
  publisher = {Springer-Verlag},
  address   = {Berlin, Heidelberg},
  pages     = {134--155},
  year      = {2024},
  doi       = {10.1007/978-3-031-65112-0_7}
}

@article{singh2026verge,
  author    = {Vikash Singh and Darion Cassel and Nathaniel Weir and Nick Feng and Sam Bayless},
  title     = {{VERGE}: Formal Refinement and Guidance Engine for Verifiable {LLM} Reasoning},
  journal   = {arXiv},
  year      = {2026},
  eprint    = {2601.20055},
  archiveprefix = {arXiv},
  doi       = {10.48550/arXiv.2601.20055}
}

@article{korner2022prolog,
  author    = {Philipp K\"{o}rner and Michael Leuschel and Jo\~{a}o Barbosa and V\'{i}tor Santos Costa and Ver\'{o}nica Dahl and Manuel V. Hermenegildo and Jose F. Morales and Jan Wielemaker and Daniel Diaz and Salvador Abreu and Giovanni Ciatto},
  title     = {Fifty Years of {Prolog} and Beyond},
  journal   = {arXiv},
  year      = {2022},
  eprint    = {2201.10816},
  archiveprefix = {arXiv},
  doi       = {10.48550/arXiv.2201.10816}
}

@article{muggleton2015meta,
  author    = {Stephen H. Muggleton and Dianhuan Lin and Alireza Tamaddoni-Nezhad},
  title     = {Meta-Interpretive Learning of Higher-Order Dyadic Datalog: Predicate Invention Revisited},
  journal   = {Machine Learning},
  volume    = {100},
  number    = {1},
  pages     = {49--73},
  year      = {2015},
  month     = jul,
  doi       = {10.1007/s10994-014-5471-y}
}

@article{yang2026hypothesis,
  author    = {Yang Yang and Jiemin Wu and Yutao Yue},
  title     = {Hypothesis Generation via {LLM}-Automated Language Bias for {ILP}},
  journal   = {arXiv},
  year      = {2026},
  eprint    = {2505.21486},
  archiveprefix = {arXiv},
  doi       = {10.48550/arXiv.2505.21486}
}

@article{pan2023logiclm,
  author    = {Liangming Pan and Alon Albalak and Xinyi Wang and William Yang Wang},
  title     = {{Logic-LM}: Empowering Large Language Models with Symbolic Solvers for Faithful Logical Reasoning},
  journal   = {arXiv},
  year      = {2023},
  eprint    = {2305.12295},
  archiveprefix = {arXiv},
  doi       = {10.48550/arXiv.2305.12295}
}

@article{he2025reasoning,
  author    = {Kaiyu He and Zhiyu Chen},
  title     = {From Reasoning to Learning: A Survey on Hypothesis Discovery and Rule Learning with Large Language Models},
  journal   = {arXiv},
  year      = {2025},
  eprint    = {2505.21935},
  archiveprefix = {arXiv},
  doi       = {10.48550/arXiv.2505.21935}
}

@misc{cattral2002poker,
  author    = {F. O. {Robert Cattral}},
  title     = {Poker Hand},
  year      = {2002},
  publisher = {UCI Machine Learning Repository},
  doi       = {10.24432/C5KW38}
}

@article{michalski1983theory,
  author    = {Ryszard S. Michalski},
  title     = {A Theory and Methodology of Inductive Learning},
  journal   = {Artificial Intelligence},
  volume    = {20},
  number    = {2},
  pages     = {111--161},
  year      = {1983},
  month     = feb,
  doi       = {10.1016/0004-3702(83)90016-4}
}

@misc{aleph2026,
  title     = {The {Aleph} Manual},
  year      = {2026},
  note      = {Accessed: Apr.\ 01, 2026},
  url       = {https://www.cs.ox.ac.uk/activities/programinduction/Aleph/aleph.html}
}
}

\end{document}